\documentclass[12pt]{JHEP3}

\usepackage{epsfig}
\epsfclipon
\usepackage{multicol}
\usepackage{epsfig,bm}
\usepackage{amssymb,amsmath}
\usepackage{graphicx}

\newcommand{\be}{\begin{equation}}
\newcommand{\ee}{\end{equation}}
\newcommand{\bea}{\begin{eqnarray}}
\newcommand{\beas}{\begin{eqnarray*}}
\newcommand{\eea}{\end{eqnarray}}
\newcommand{\eeas}{\end{eqnarray*}}
\newcommand{\ba}{\begin{array}}
\newcommand{\ea}{\end{array}}
\def\ls{\mathrel{\lower4pt\vbox{\lineskip=0pt\baselineskip=0pt
           \hbox{$<$}\hbox{$\sim$}}}}
\def\gs{\mathrel{\lower4pt\vbox{\lineskip=0pt\baselineskip=0pt
           \hbox{$>$}\hbox{$\sim$}}}}

\newcommand{\e}{\textrm{e}}

\newcommand{\N}{\widetilde{N}}

\def\smiley{\hbox{\large$\bigcirc$\hspace{-.80em}%
\raise.2ex\hbox{$\cdot\cdot$}\kern-.61em    
\lower.2ex\hbox{\scriptsize$\smile$}}\ }

\newcommand{\roughly}[1]{\mathrel{\raise.3ex\hbox{$#1$\kern-0.85em
\lower1ex\hbox{$\sim$}}}}

\newcommand{\lsim}{\roughly<}
\newcommand{\gsim}{\roughly>}

\def\be{\begin{equation}}
\def\beq\begin{equation}
\def\ee{\end{equation}}
\def\bea{\begin{eqnarray}}
\def\eea{\end{eqnarray}}

\def\beq{\begin{equation}}
\def\eeq{\end{equation}}
\def\beqa{\begin{eqnarray}}
\def\eeqa{\end{eqnarray}}

\def\cT{{\cal T}}

\newcommand{\bmat}{\left(\begin{array}}
\newcommand{\emat}{\end{array}\right)}

\title{Multiple Inflation, Cosmic String Networks and the String Landscape}

\author{C.P. Burgess,$^{1,2,3}$ Richard Easther,${}^4$
 Anupam Mazumdar,$^{5}$
 David F. Mota$^{6,7}$ and Tuomas Multam\"aki$^{5}$\\
 ${}^1$ Physics Department, McGill University,
 Montr{\'e}al, QC, Canada, H3A 2T8.\\
 ${}^2$ Department of Physics and Astronomy, McMaster
 University,\\
 ~~~~~Hamilton, ON, Canada, L8S 4M1. \\
 ${}^3$ Perimeter Institute, Waterloo, ON, Canada, N2L 2Y5.\\
 ${}^4$ Department of Physics, Yale University, New Haven CT, USA
 06520.\\
 $^{5}$~NORDITA, Blegdamsvej-17, Copenhagen, DK-2100, Denmark.\\
 ${}^6$ Institute of Theoretical Astrophysics,
        University of Oslo,
        0315 Oslo, Norway\\
${}^7$ Department of Physics, University of Oxford, Oxford, OX1 3RH, UK}

\date{}

\abstract{Motivated by the string landscape we
examine scenarios for which inflation is a two-step process, with
a comparatively short inflationary epoch near the string scale and
a longer period at a much lower energy (like the TeV scale). We
quantify the number of $e$-foldings of inflation which are
required to yield successful inflation within
this picture. The constraints are very sensitive to
the equation of state during the epoch between the two
inflationary periods, as the extra-horizon modes can
come back inside the horizon and become reprocessed. We find that
the number of $e$-foldings during the first inflationary epoch can
be as small as 12, but only if the inter-inflationary period is
dominated by a network of cosmic strings (such as might be
produced if the initial inflationary period is due to the
brane-antibrane mechanism). In this case a further 20 $e$-foldings
of inflation would be required at lower energies to solve the late
universe's flatness and horizon problems.}

\preprint{McGill-04/27~NORDITA-2005-02.}


\begin{document}


\section{Introduction and Summary}

The search for inflationary solutions within string theory is
still young, but considerable progress has been made over the past
several years. Most of this recent progress began with the
recognition that the separation of two branes could function as an
inflaton \cite{DvaliTye}, and that configurations of branes and
antibranes together break supersymmetry in a calculable way,
potentially allowing the explicit computation of the resulting
inflaton
dynamics\cite{BI,BI1,Angles,Renata,Others,Racetrack}.\footnote{For
recent reviews with more complete references, see
ref.~\cite{BIReviews}.} Direct contact with string theory also
became possible \cite{KKetalT} with the recognition that moduli
stabilization can be achieved at the string scale within a
controllable approximation \cite{GKP,Sethi}.

Despite it still being a young topic, it is possible to draw some
preliminary conclusions about the implications of embedding
inflationary cosmology into  string theory. In particular, the
following two features are seen in  a large set of models
\cite{RealInfl}.

\begin{itemize}
\item It is no easier  to obtain 60 $e$-foldings of inflation
within string theory  than in other models. For all of the known
examples of inflation within string theory there are parameters
and/or initial conditions which must be carefully tuned. The
problem arises because inflation typically occurs at energies near
the string scale, and so there are no particularly small
dimensionless numbers from which to obtain inflation. Of course,
the fewer $e$-foldings which are demanded, the less precise are
the adjustments that are required.
\item Although the most recent progress has focused on Type IIB
vacua having few moduli, supersymmetric string vacua are usually
littered with moduli, which generically acquire masses only after
supersymmetry breaks. Thus we generically find numerous scalar
fields whose masses are comparatively small, since they are close
to the weak scale, $M_w$. Scalars of this type are usually
anathema for cosmology since they cause a host of problems for
cosmology \cite{moduliproblems} during the Hot Big Bang.
\end{itemize}

It has been suggested that these two problems may be each other's
solutions \cite{RealInfl,Folded}. For instance, suppose many fewer
than 60 $e$-foldings of inflation occur up at the string scale,
whose quantum fluctuations generate the large-scale structure
which is presently visible as temperature fluctuations in the
Cosmic Microwave Background (CMB). In this case the remaining
$e$-foldings of inflation  required by the horizon and flatness
problems might plausibly be obtained during later epochs, when the
many moduli are rolling around in the complicated potential
landscape which is generated by low-energy supersymmetry breaking.
This later period of inflation might be due to a mechanism which
depends on the more complicated potentials expected in a
low-energy theory involving very many would-be moduli, perhaps
along the lines proposed in ref.~\cite{dk,Easther:2004qs}. Besides
solving the horizon and flatness problems, if it occurs at the
weak scale this second stage of inflation could also solve some of
the problems associated with moduli along the lines suggested in
ref.~\cite{lyth-stewart}.

Thus, it may be that a more generic picture of inflation within
string theory is a two-step (or multi-step) process --- as in
Fig.~\ref{fig:fig0} --- given what we know about the low-energy
potential landscape. If so, then there would be several
observational consequences.

\begin{enumerate}
\item The observed CMB temperature fluctuations should not be deep
within the slow-roll regime, since $N_e < 60$ implies the slow
roll parameters are likely to be larger: $\eta \sim 1/N_e > 0.02$.
In particular perhaps the slow-roll parameters might instead be of
order $1/N_e \sim 0.1$ or larger. If so, then phenomena like kinks
in the inflationary spectrum or a running spectral index would be
much more likely to arise at observable levels. This realizes the
folded inflation scenario \cite{Folded}, albeit with a long hiatus
between inflation in the different directions in the landscape.
This model is naturally compatible with a sizable tensor
contribution to the observed spectrum of CMB anisotropies.

\item A period of late inflation would require the explanation of
any post-inflationary processes like baryogenesis to take place at
energies as low as the electroweak scale, with all of the
associated observational implications which this would imply.
\end{enumerate}


\FIGURE[ht]{\epsfig{file=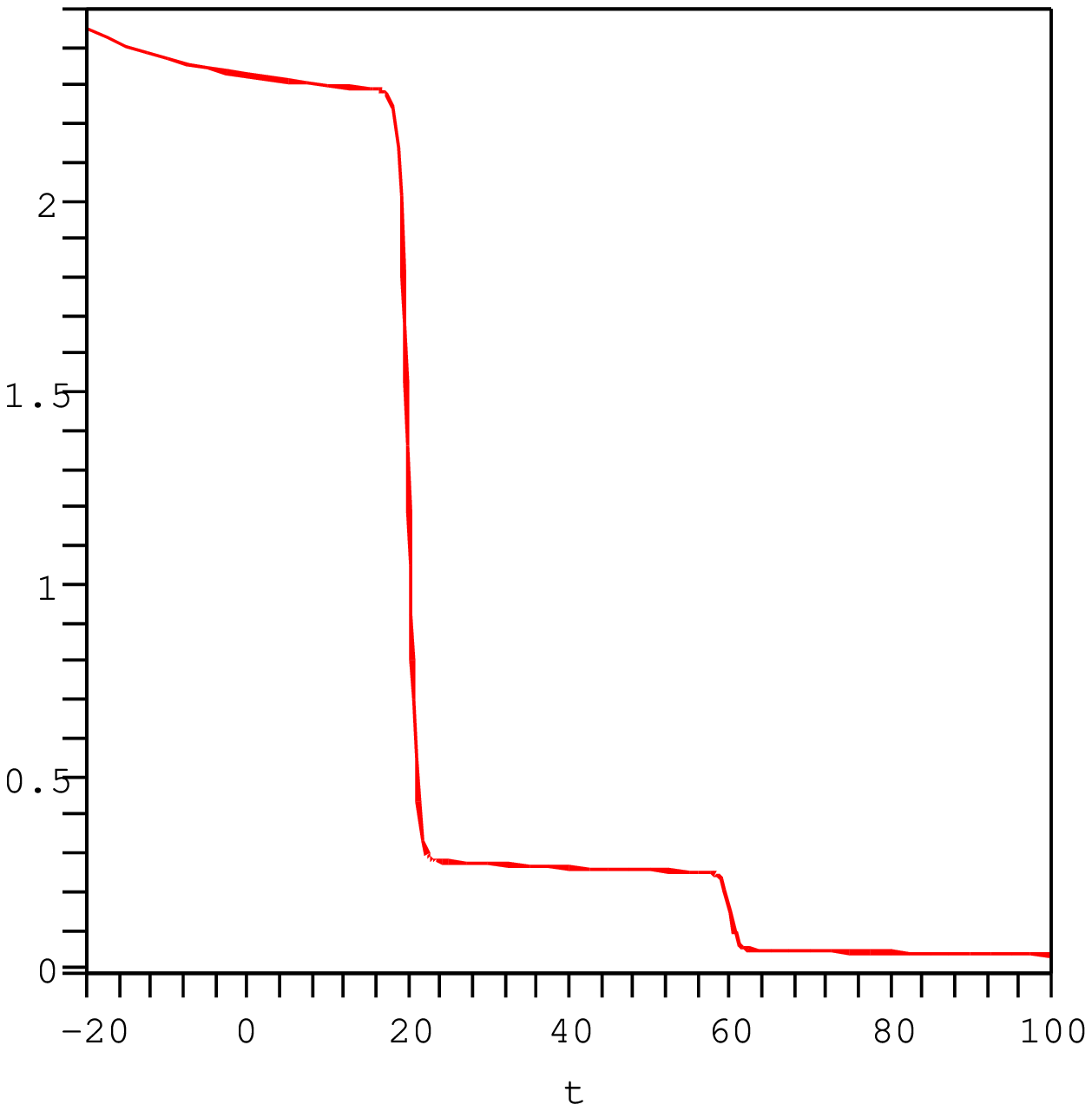,width=10cm}
  \caption{A sketch of the Hubble scale, $H$, during a two-step
  inflationary process. We imagine the first step to involve a minimal
  number of $e$-foldings at the string scale, followed by more
  $e$-foldings at the electroweak scale.}}\label{fig:fig0}

This paper examines the consequences of this kind of two-step
inflationary cosmology. Our goal is to determine the minimal
amount of inflation needed at high energies to produce the CMB
temperature fluctuations, and how this minimal amount depends on
the subsequent history of the universe.\footnote{Because our
interest is in scenarios for which the second inflationary epoch
occurs at the electroweak scale, modes which leave the horizon
during this epoch produce CMB fluctuations whose amplitudes are
too small to be visible. As such, it differs from previous
two-step inflationary proposals in the literature
\cite{PreviousTwoStep}, which typically aim to  produce an
observable feature in the present-day large-scale structure.} We
employ general arguments based on energy conservation, and
construct explicit toy models for inflationary potentials which
exhibit multiple inflationary periods.

We note in passing that while the kind of two-stage inflation
envisioned here may well involve less fine-tuning than a single,
longer stage of inflation at higher energies.  However, very
unusual inflationary trajectories may still be ``preferred'' if
they lead to vast amounts of inflation, as the low probability of
such a hypothetical trajectory may be offset by the vast volume of
space it produces \cite{multiverse}.

The remainder of the paper is organized as follows. \S2, which
follows immediately, describes the bounds on the amounts of
inflation which can be derived using only general arguments about
the size and equation of state of the universe. This is followed,
in \S3, by a description of a toy field theory which has been
designed to produce a simple concrete example having multiple
inflationary epochs. In an appendix we present the most general
set-up of multiple inflation and the total number of $e$-foldings
required to solve the horizon problem. Our conclusions are
summarized in \S4.

\section{General Constraints}

Here we determine some general constraints
on the number of $e$-foldings need in any two-step inflationary
model. Our analysis follows closely that of
ref.~\cite{liddle}. The essential features of the inflationary
process we assume are captured by Fig.~\ref{fig:fig0}. This figure
shows the Hubble scale over two inflationary epochs, followed by
power-law evolution for the  scale factor.

The  constraints quantified in this section are of three
types. First, we must ask for the total amount of expansion to be
large enough to solve the horizon and flatness problems in the way
which motivated inflationary scenarios in the first place. Second,
we demand that the first inflationary phase lasts long enough to
generate all of the primordial fluctuations which are observed on
long scales during the present epoch. Finally, we require that
none of the modes leaving the horizon during the first epoch of
inflation are reprocessed in an observable way during the
subsequent  evolution of the universe.


\FIGURE{\epsfig{file=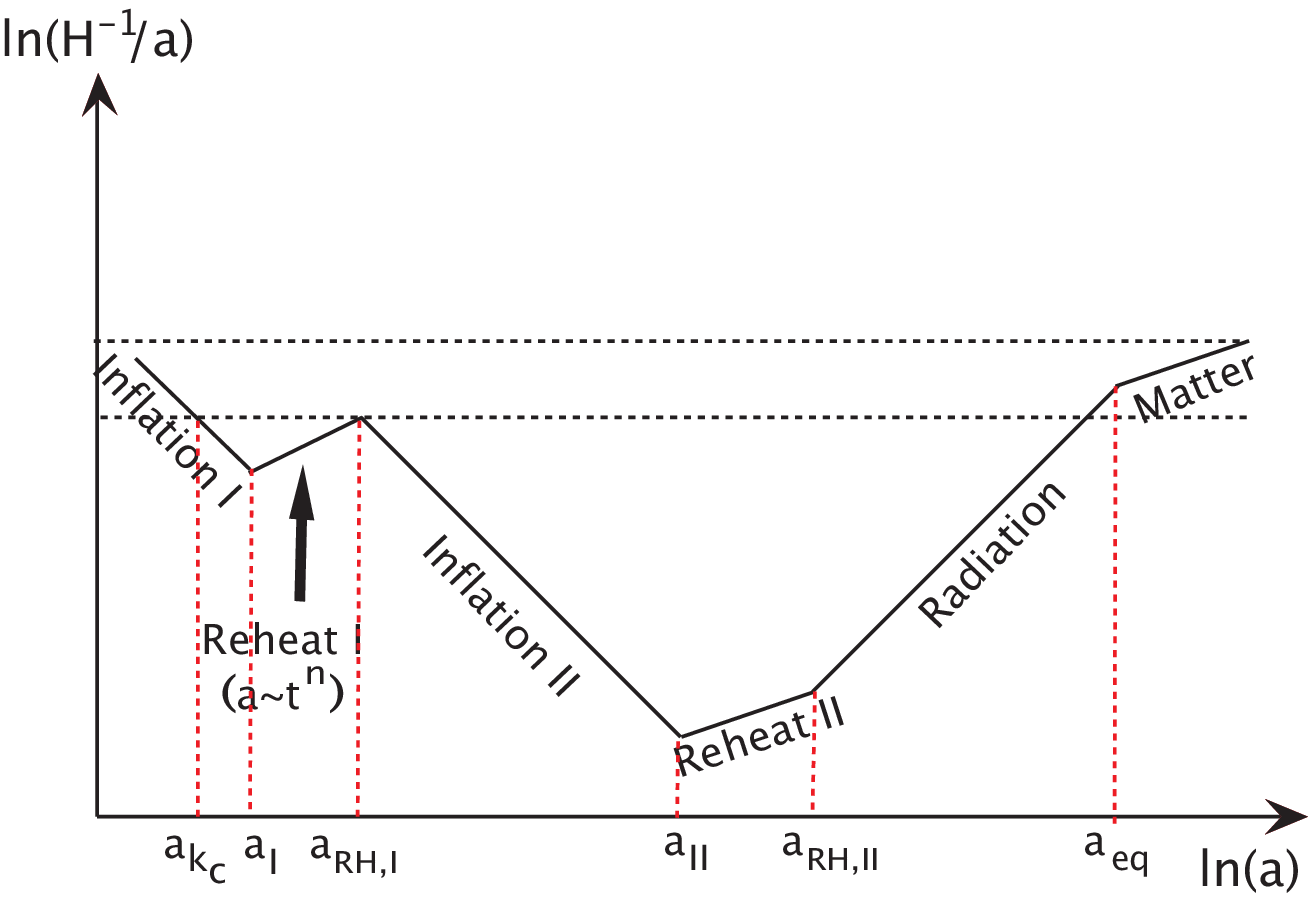,width=10cm}
  \caption{A sketch of the comoving Hubble scale during a two-step
  inflationary process. We plotted $\ln(H^{-1}/a)$ versus $\ln(a)$.
  During inflation by definition the comoving Hubble length decreases,
  otherwise it increases as it is shown in the figure.  We
  parameterized the phase between the two inflationary epochs by a
  generic power law behavior with $n\leq 1$. For $n=2/3$ the Universe
  will be dominated by the matter dominated epoch, for $n=1/2$ the
  Universe is filled with radiation and for $n=1$ the expansion is
  dominated by the network of strings.}}\label{fig:fig1}

\subsection{Cosmological Kinematics}

In order to derive the constraints implied by these conditions, it
is convenient to plot the quantity $\ell = 1/(aH)$ against $a$,
where $a$ is the  scale factor and $H = da/dt$, as in
Fig.~\ref{fig:fig1}. Inflationary epochs appear on such a plot as
regions for which $\ell$ shrinks as $a$ increases, while for
ordinary matter- or radiation-dominated epochs $\ell$ is a
growing function of $a$. Since scales leave the horizon when their
co-moving wave-number, $k$, satisfies $k/a = H$, the epochs when a
particular mode is inside (or outside) of the horizon corresponds
to the times when the curve for $y = \ell$ falls below (or above)
a horizontal line drawn at $y = 1/k$. For instance, the uppermost
of the horizontal dotted lines in Fig.~\ref{fig:fig1} corresponds
to the scale, $k_0$, which is just re-entering the horizon at the
present epoch.

As the picture makes clear,  the last modes to leave the horizon
during the first epoch of inflation subsequently re-enter the
horizon during the interval before the second inflationary period.
The lower line in Fig.~\ref{fig:fig1} represents the critical
scale, $k_c$, which is the last scale to re-enter in this way
before the second inflationary epoch. The standard inflationary
calculations for the density fluctuations produced by a mode
assume that it does not get reprocessed in this way by the
intermediate evolution between its initial horizon exit and its
re-entry inside the horizon during the present epoch, and so in
what follows we simplify the problem by demanding that none of the
modes which are currently observable get reprocessed in this way.

Our goal is to quantify the conditions which must be satisfied by
the number of $e$-foldings, $N_I$ and $N_{II}$, which occur during
the two inflationary periods, defined as follows:
\be \label{efolddefs}
    \e^{N_I(k)}\equiv {a_I\over a_k},\ \qquad
    \e^{N_{II}}\equiv {a_{II}\over a_{RH,I}} \,.
\ee
Here $a_{I}$ (and $a_{II}$) is the value of the scale factor at
the end of first (and second) phase of inflation, while $a_{RH,I}$
similarly denotes the scale factor at the end of the `reheating'
period which occurs in between the two bouts of inflation.

Of particular interest is how short it is possible to make the
initial inflationary period, $N_I$. For this reason it is also
convenient to define the number, $\N$, of $e$-foldings of
expansion between the time the critical scale $k_c$ leaves the
horizon until the end of first inflationary period:
\be \label{efolddefs2}
    \e^{\N}\equiv {a_I\over a_{k_c}} \,.
\ee
Since $N_I > \N$, it is important to know how large $\N$ must be,
for various choices for the intermediate inflationary physics.

Following \cite{liddle}, we may derive a relationship amongst the
quantities $N_I$, $N_{II}$ and $\N$, by expressing the condition
that a particular mode, $k$, which enters the horizon at late
times also left the horizon during the first inflationary epoch.
That is, if a mode, $k$, leaves the horizon when the scale factor
and Hubble scale had values $a_k$ and $H_k$, then the defining
condition $k = a_k H_k$ implies:
\bea  \nonumber
    {k\over a_0H_0} & = & {a_kH_k\over a_0H_0} \\
    & = & \left( {a_k\over a_I} \right) \left( {a_I\over a_{RH,I}}
    \right) \left( {a_{RH,I}\over a_{II}} \right)
    \left( {a_{II}\over a_{RH,II}} \right) \left(
    {a_{RH,II}\over a_{eq}} \right) \left(
    {H_k\over H_{eq}} \right) \left( {a_{eq}H_{eq} \over
    a_0H_0}\right) \nonumber \\
    & = & \e^{-N_I(k)} \left( {a_I\over a_{RH,I}} \right)
    \e^{-N_{II}} \left( {a_{II}\over a_{RH,II}} \right)
    \left( {\rho_{RH,II} \over \rho_{eq}} \right)^{-1/4}
    \left( {H_k\over H_{eq}} \right) \left( {a_{eq}H_{eq}
    \over a_0 H_0} \right). \label{efolds}
\eea
To obtain the final line above, we used the fact that $\rho_r
\propto a^{-4}$ to replace ${a_{RH,II}\over a_{eq}}$, along with
the  Hubble scale at the present time, $H_0$, and at
radiation-matter equality, $H_{eq}$.  In what follows we take $H_0
= h \, (100 \, \hbox{km/s/Mpc}) = 1.75 \times 10^{-61}\, h\, M_p$,
where $h \approx 0.7$ and $M_p = G^{-1/2}$ is the Planck mass (in
units for which $\hbar = c = 1$). Similarly, $H_{eq}=5.25 \times
10^6\, h^3\, \Omega_m^2\, H_0$ \cite{liddlebook}, where $\Omega_m
\approx 0.25$ is the present matter density in units of the
critical density. Finally, the ratio of the product $aH$ at
present and at radiation-matter equality is known to be
$a_{eq}H_{eq}/a_0 H_0 \approx 219 \, \Omega_m \, h$
\cite{liddlebook}.

During the inflationary periods we assume the energy density to be
approximately constant, $\rho_I \approx \rho_k \approx \rho_{k_c}
\approx V_I$, leading also to a constant Hubble scale $H_I \approx
H_k \approx H_{k_c} \approx \left( {8\pi V_{I}/3 M_p^2}
\right)^{1/2}$. Similarly $\rho_{RH,I} \approx \rho_{II} \approx
V_{II}$ and $H_{RH,I} \approx H_{II} \approx \left( 8 \pi V_{II}/3
M_p^2 \right)^{1/2}$.

Finally, we must assume an equation of state during the two
reheating epochs. For convenience we parameterize the
expansion rate during these epochs by a power law, $a(t) \propto
t^n$, which is equivalent to $\rho(a) \propto a^{-r}$ with $r =
2/n$, or to a constant equation-of-state parameter $w = p/\rho =
(2/3n) - 1$. Some common values of all three parameters are listed
in Table 1.

\TABLE[ht]{
\centerline{
\begin{tabular}{||c||c|c|c|c|c||}
\hline & Cosm. Const.  & String Network & Matter & Radiation &
Kinetic Dom. \\
\hline $w$ & $-1$  & $-1/3$ & 0 & 1/3 & 1 \\
$r$ & 0 & 2 & 3 & 4 & 6 \\
$n$ & $\infty$ & 1 & 2/3 & 1/2 & 1/3 \\
\hline
\end{tabular}}
\caption{The parameters $n$ and $r$ discussed in the text for some
choices for the equation of state parameter $w = p/\rho$.}
}\label{paramtable}

The relation $(a_{II}/a_{RH,II}) =
(V_{II}/\rho_{RH,II})^{-1/r_{RH,II}}$ follows with these choices
for the second reheat epoch, where $r_{RH,II} = 2/n_{RH,II}$
parameterizes the equation of state as above. If, for instance,
this reheating occurs due to the decay of a coherently oscillating
inflaton field then $r_{RH,II} = 3$, as is appropriate for a
matter-dominated universe. Since, for numerical applications, we
typically assume instant reheating for simplicity --- {\it i.e.}
$\rho_{RH,II} \approx V_{II}$ --- our results are not sensitive to
the value of $r_{RH,II}$.

A simple expression may be derived for $\N$, as a function of $n =
n_{RH,I}$ and the energy densities during the two inflationary
epochs. Notice that during the inter-inflationary period $\rho
\propto H^2 \propto 1/t^2 \propto a^{-2/n}$ and so $\ln[1/(aH)] =
[(1-n)/n] \, \ln a$. Because this is an increasing function of $a$
provided $0< n < 1$, for $n$ in this range there are modes which
re-enter the horizon during this epoch, making $\N$ nonzero. Since
$V_I/V_{II} = \rho_{I}/\rho_{RH,I} \propto
(a_{RH,I}/a_{I})^{2/n}$, $a_{k_c} H_{k_c} = a_{RH,I} H_{RH,I}$
(see Fig.~\ref{fig:fig1}) and $H_{RH,I}/H_{k_c} = H_{II}/H_I =
(V_{II}/V_I)^{1/2}$, it follows that
\be \label{infcalc}
    {a_{RH,I} \over a_I}  = \left( {a_{RH,I} \over a_{k_c}}
    \right) \left( {a_{k_c} \over a_I} \right)
    = \left( {H_{I} \over H_{II}} \right)
    \left( {a_{k_c} \over a_I} \right)
    = \left( {V_{I} \over V_{II}} \right)^{1/2}
    \e^{-\N} \,.
\ee
But since $V_I/V_{II} = (a_{RH,I}/a_I)^{2/n}$ we may eliminate
either $V_I/V_{II}$ or $a_{RH,I}/a_I$ from this equation to obtain
\be \label{infcalc1}
    \N = \left( {1-n \over 2} \right) \,
    \ln \left( {V_I \over V_{II}} \right)
    = \left( {1-n \over n} \right) \,
    \ln \left( {a_{RH,I} \over a_{I}} \right)
     \,.
\ee
As expected, this is positive for $0 < n < 1$, and vanishes in the
special case $n=1$, since in this case $aH$ is a constant during
the inter-inflationary period (corresponding to a horizontal line
in Fig. \ref{fig:fig1}). For example, choosing $V_I^{1/4} =
10^{16}$ GeV, $V_{II}^{1/4} = 10^{3}$ GeV and $n=2/3$ (as for
matter domination) implies $\N\approx 20$.

Substituting eq.~(\ref{infcalc}) into eq.~(\ref{efolds}), taking a
logarithm and rearranging gives
\bea \label{lnefolds}
    N_I(k)+{n \N \over 1-n} + N_{II} & = &
    -\ln\left({k\over a_0H_0}\right)+ \frac{1}{r_{RH,II}}
    \ln\left({\rho_{RH,II}\over
    V_{II}}\right)+
    \frac 14 \ln\left({\rho_{eq}\over
    \rho_{RH,II}}\right)\nonumber\\
    & &\qquad  +\ln\left( {1\over H_{eq}}
    {\sqrt{8\pi V_k\over 3 M_p^2}
    }\right)+
    \ln(219\, \Omega_m h).
\eea
Further simplifying by assuming instant reheating after the second
phase of inflation ({\it i.e.} $\rho_{RH,II} = V_{II}$),
specializing to the mode, $k_0 = a_0 H_0$, which is presently
entering the horizon and using the numerical values for quantities
at radiation-matter equality gives:
\be \label{lnefolds2}
    N_I + {n \N \over 1-n}  + N_{II} = 68.5 +
    \frac14 \ln \left( {V^2_{I} \over V_{II} M_p^4}
    \right),
\ee
where we define $N_I \equiv N_I(k_0)$. Finally, making use of
eq.~(\ref{infcalc1}) to eliminate $\N$ gives
\be \label{lnefolds3}
    N_I + N_{II} = 68.5 +
    \frac14 \ln \left[ {V_{II} \over M_p^4} \left(
    { V_I \over V_{II}} \right)^{2(1-n)}
    \right] \,.
\ee
This, together with eq.~(\ref{infcalc1}), is one of the main
results of this section.

\subsection{Constraints}

We are now in a position to quantify the circumstances which
minimize the lengths of the two inflationary epochs. There are
three conditions which the parameters $N_I$ and $N_{II}$ must
satisfy.

\begin{enumerate}
\item We require that the initial inflationary epoch generate the
large-scale temperature fluctuations in the Cosmic Microwave
Background (CMB) through the usual mechanism. Since the density
contrast, $\delta$, generated in this way is related to the
inflationary energy density, $V_I$, and slow-roll parameter
$\epsilon = \frac{1}{16 \pi} \, (M_p V'/V)^2$ by \cite{liddlebook}
\be \label{deltavsVI}
    \delta^2 = \frac{32}{75} \, \left( \frac{V_I}{\epsilon
    \, M_p^4} \right) \,,
\ee
agreement with observations implies $\delta \approx 10^{-5}$ and
so $V_I/M_p^4 \approx 10^{-7} \, \epsilon$.

\item We require  the initial inflationary epoch to last long
enough so that its termination does not introduce any features
into the observed large-scale structure. The current horizon is of
order $H_0^{-1} \sim 10$ Gpc. On the other hand, those modes on
the order of 10 Mpc are starting to feel nonlinear effects, and it
is thus difficult to  constrain primordial density fluctuations on
smaller scales.  We thus require that the initial inflationary
epoch last long enough to generate a scale-invariant spectrum over
scales differing by a factor of $10^3$, and that these modes not
be reprocessed by the subsequent inter-inflationary evolution.
This corresponds to the requirement that there be at least
\be
    N_I > \N + \ln \left( 10^3 \right) \approx 6.9 +
    \left( {1-n \over 2} \right) \,
    \ln \left( {V_I \over V_{II}} \right)
\ee
$e$-foldings during the initial inflation. For example, taking as
before $V_{II}^{1/4} = 10^{3}$ GeV, $V_I^{1/4} = 10^{16}$ GeV and
assuming matter domination during the inter-inflationary epoch
then implies $N_I \gsim 27$.
\item Finally, eq.~(\ref{lnefolds3}) states how much inflation
there must be in total in order to solve the horizon and flatness
problems, as functions of the energies during the inflationary
epochs and the equation of state during the inter-inflationary
period. Using eq.~(\ref{deltavsVI}) to eliminate $V_I$ from this
equation gives the final result
\be \label{lnefolds4}
    N_I + N_{II} = 68.5 + \left( \frac{1-n}{2} \right)
    \ln \left( {10^7 \over \epsilon}
    \right) + \left( \frac{1 - 2n}{4} \right) \ln \left(
    {M_p^4 \over V_{II}} \right) \,.
\ee

\end{enumerate}

We see that an absolute lower limit to the number of $e$-foldings
of inflation which is responsible for seeding the observed
large-scale structure is $N_I \approx 7$, and this rock-bottom
value is only possible if either $n = 1$ or $V_{II} = V_I$. If one
attempts to achieve $N_I \approx 7$ by choosing $V_{II} = V_I$
then the universe does not expand at all between the inflationary
periods, and this reduces to the situation of a single
inflationary epoch. In this case the total duration of inflation
is $N_{tot} = N_I + N_{II}$, which eq.~(\ref{lnefolds3}) shows
must satisfy
\be \label{lnefolds3a}
    N_{tot} = 68.5 +
    \frac14 \ln \left( {V_{II} \over M_p^4} \right)
    = 64.5 + \frac14 \ln \epsilon \,.
\ee
Clearly $N_{tot}$ is then minimized in the usual way, by choosing
$V_{II} = V_I$ as small as possible. For instance, in the extreme
situation where $V_{II} \sim (10^3 \; \hbox{GeV})^4$ we have
$N_{tot} \approx 31.7$. Such a scenario also requires an extremely
slow roll, since eq.~(\ref{deltavsVI}) implies $\epsilon \sim 7
\times 10^6 (V_{II}/M_p^4) \sim 10^{-57}$.

A more interesting scenario arises if $n=1$, since in this case no
modes ever re-enter the horizon during the inter-inflationary
period. Consequently, at face value, $\N = 0$ and so $N_I$ can be
as small as 7 without causing direct problems for CMB
observations. We now examine this situation in somewhat more
detail. (In the appendix we provide a generalization of the
two-fold inflationary results,
eqs.~(\ref{lnefolds3},\ref{lnefolds4}), to the case where there
are multiple bouts of inflation, separated by intervening
inter-inflationary epochs.)

\section{The Inter-Inflationary String Network}

The previous section shows that if $n = 1$ (or $w = -1/3$) during
the inter-inflationary phase we are actually on the borderline
between regular and inflationary expansion, and the the quantity
$aH$ is approximately constant. As a result, modes neither exit or
re-enter the horizon.  This kind of equation of state can arise if
the energy density of the universe is dominated by a network of
cosmic strings. We now ask in a bit more detail what would be
required for the success of such a scenario.

In this case the constraints on the amount of inflation due to
horizon and flatness problems may be solved simply by
adjusting $N_{II}$ to satisfy eq.~(\ref{lnefolds4}), which may be
written in the form
\be
    N_I + N_{II} = 68.5 -  \frac{1}{4} \, \ln \left(
    {M_p^4 \over V_{II}} \right) \,.
\ee
As we found previously for the case $V_I = V_{II}$, this can allow
$N_I + N_{II}$ to be as small as 32 if $V_{II}$ is as low as
$10^3$ GeV. An important difference between the alternatives $n=1$
and $V_I = V_{II}$, however, is that if $n=1$ the choice $V_{II} =
10^3$ GeV does not automatically imply an incredibly small value
for the slow-roll parameter, $\epsilon$, during the first
inflationary phase in order to explain the observed CMB
fluctuation spectrum.

\subsection{Mode Reprocessing}

When $n=1$, modes leaving the horizon during the first inflationary
period automatically remain outside the horizon during the
inter-inflationary phase. At face value there would appear no
additional constraints on the inter-inflationary phase coming from the
requirement that these modes not be reprocessed between the two
inflationary epochs. However, in this scenario the modes which are
just outside the horizon remain just outside the horizon for this
entire epoch, which can be for a very long time if $V_{I} \gg
V_{II}$. We now ask in more detail how the proximity of these modes to
the horizon for such a long period might affect the density
perturbations to which they ultimately give rise.

We examine this issue within the relatively simple context where
the energy density during the inter-inflationary period is
dominated by the evolution of a scalar field, $\phi$. In this case
the evolution of the curvature perturbation $\mathcal{R}$ is
governed by the equation \cite{mfb}
\be \label{uequ}
    \partial_\tau^2u_k + \left( k^2 - {\partial_\tau^2 z\over z}
    \right)u_k=0,
\ee
where $u_k$ is the Fourier component of the gauge-invariant
potential, $u_k=|z \mathcal{R}_k|$, $z \equiv a\dot{\phi}/H$ and
$\tau$ denotes conformal time.

This equation admits analytic solutions during a period when the
rolling of the background field $\phi$ mimics a constant equation
of state. That is, if $p = \frac12 \dot\phi^2 - V(\phi)$ and $\rho
= \frac12 \dot \phi^2 + V(\phi)$, then if $p =w \rho$ for constant
$w$ it is straightforward to see that eq.~(\ref{uequ}) can be
written
\be \label{uequ2}
    u_k''-\frac12 (1+3w)u_k' + \left[ \left({k\over H_0} \right)^2
    a^{(1+3w)}-\frac 12(1-3w)\right] u_k = 0,
\ee
where the primes denote differentiation with respect to $\ln a$
and $H_0$ denotes $H$ at an initial epoch, for which $a = a_0$.
Notice that this equation is {\it not} valid for the special case
$w=-1$, which is the case of a pure cosmological constant, since
in this case the quantity $z$ vanishes. The growing solution for
this equation is solved in terms of Bessel functions,
\be \label{usol}
    u_k(a) = u_{k0} \; a^{\frac 14 (1+3w)} H^{(1)}_\nu
    \left[ {2k \, a^{\frac 12 (1+3w)}\over H_0 (1+3w)}  \right],
\ee
where
\be
    \nu = {3(1-w) \over 2(1+3w)} \,,
\ee
and the normalization constant, $u_{k0}$, is left arbitrary.

Because of the singularities in these expressions for $w = - 1/3$,
this case must be treated separately. Specializing the starting
point, eq.~(\ref{uequ2}), to $w = -1/3$ gives the simple form
\be \label{uequ3}
    u_k''+ \Big( k^2 - 1 \Big) u_k =0,
\ee
where for simplicity we choose units such that $a_0 H_0 = 1$. It
is trivial to integrate this equation and see that the
super-horizon modes (for which $k<1$) behave like
\be
    u_k \propto a^{\pm\sqrt{1-k^2}} \,.
\ee
Our assumption that $w$ is approximately constant implies the same
also holds for $\dot\phi$, and so if $w = -1/3$ then $H \propto
1/a$ further implies $z \propto a^2$. We therefore see that the
curvature perturbations on super-horizon scales evolve during the
inter-inflationary phase according to
\be \label{superh}
    \mathcal{R}_k(a) = \mathcal{R}_k(a_0) \, \left(
    {a \over a_0} \right)^{\sqrt{1-k^2}-2} \,.
\ee

This expression shows how density fluctuations hovering near the
horizon get reprocessed during the inter-inflationary phase. The
resulting power spectrum is
\be \label{power}
    \mathcal{P}_\mathcal{R}(k) = \left( {k^{3} \over 2\pi} \right)
    \left| \mathcal{R}_k \right|^2  \,,
\ee
and so as the universe expands from $a_I$ to $a_{RH,I}$ between
inflationary epochs the power spectrum is modified according to
\be \label{power2}
    \mathcal{P}_\mathcal{R} \left( k,a_{RH,I} \right) =
    \mathcal{P}_\mathcal{R} \left( k,a_{I} \right)
    \left( {a_{RH,I} \over a_I} \right)^{2\sqrt{1-k^2} - 4}.
\ee

We see that modes just outside the horizon (near $k = 1$) {\it
are} reprocessed, and this superimposes a $k$-dependence onto the
power spectrum generated during the first inflationary epoch. The
corresponding spectral index is $n_s -1 = d \ln
\mathcal{P}_{\mathcal{R}} /d \ln k$, and so
\bea \label{specindexmod}
    n_s -1 &=& (n_s - 1)_{\rm inf} - { 2 k^2 \over
    \sqrt{1 - k^2}} \, \ln \left(
    {a_{RH,I} \over a_I} \right) \nonumber \\
    &=& (n_s - 1)_{\rm inf} - { k^2 \over \sqrt{1 - k^2}}
    \, \ln \left(
    {V_{I} \over V_{II}} \right)\,,
\eea
where the first term represents the contribution due to the
inflationary phase and the second equality uses
eq.~(\ref{infcalc1}) to express the amount of  expansion
between the inflationary epochs to the relative energy densities
when inflation is taking place. This introduces a strong
$k$-dependence for those modes with $k$ near unity, which
corresponds to those which hover right on the horizon.

If we demand that the modes which are currently observable have
not been unacceptably processed in this way, then we may derive a
minimal amount of $e$-foldings which must take place after the
last observable mode leaves the horizon and before the onset of
the inter-inflationary phase. Taking $V_I^{1/4} \sim 10^{16}$ GeV
and $V_{II}^{1/4} \sim 10^3$ GeV, we see that $n_s - n_{s,{\rm
inf}} \approx - 120 k^2/\sqrt{1-k^2}$, and so demanding this
quantity to be smaller than order 1\% implies $k \lsim 0.01$.

Finally, we determine the number of $e$-foldings during the first
inflationary phase between the exit of modes having $k \sim 0.01$
and those having $k = 1$, since this gives the minimal amount of
inflation which must occur after the exit of the last observable
mode before the inter-inflationary period can begin. Since $aH$
must increase by a factor of 100 during this period, it involves a
total of $\tilde N \sim \ln 100 \sim 5$ $e$-foldings. If fewer
than this number of $e$-foldings occurs between the exit of
observable modes and the end of the first inflationary period,
then the processing of these modes during the inter-inflationary
period would become unacceptably large.

These 5 $e$-foldings must be added to the roughly 7 $e$-foldings
which must occur while the observable scales exit the horizon,
leading us to a bare minimum of $N_I \gsim 12$ $e$-foldings during
the first inflationary epoch if we are not to ruin the success of
the inflationary predictions for the CMB temperature fluctuations.
As we have seen, this must be followed by at least another 20
$e$-foldings of inflation during the second inflationary epoch in
order to successfully solve the late-time flatness and horizon
problems.

It would clearly be of considerable interest to more carefully
compute the mode reprocessing which can occur during the
inter-inflationary phase in order to go beyond the simplistic
scalar-field model of this phase which is used here.

\subsection{Obtaining a String Network}

Can an inter-inflationary epoch with $n=1$ can actually arise
within our current understanding of inflation? Since $n=1$
corresponds to an equation of state $w=-1/3$, it is what would be
expected if a network of cosmic strings were to dominate the
universe immediately after the first inflationary epoch. But why
wouldn't the first inflationary phase simply inflate away any
cosmic strings which might be present?

Interestingly, it appears that post-inflationary cosmic strings
can arise naturally within the framework of brane-antibrane
inflation, as cosmic strings are generically generated by the
brane-antibrane annihilation which marks the end of the
inflationary phase \cite{BI,cosmicstrings,stringycosmicstrings}.
We believe this to be a very intriguing possibility whose
implications are worth studying in more detail.

The resulting cosmic string network must have several properties
in order for this kind of cosmology to emerge.\footnote{We thank
Robert Brandenberger for many useful suggestions on this topic.}
First, its energy density must dominate the universe after the
first inflationary epoch ends. Second, the string networks must be
sufficiently long-lived to allow them to dominate the energy
density for an appreciable period of time. Neither of these
properties is generically true for strings which form as defects
during phase transitions in weakly-coupled gauge theories
\cite{stringnetwork}, as we now see.

What is expected for the initial energy density of a string
network within a perturbative gauge theory is most easily
estimated within a toy model, like the electrodynamics of a
charged scalar $\phi$. The crucial quantity which determines the
properties of the string network is the correlation length, $\xi$,
for the phase of $\phi$ at the epoch when the temperature falls to
the temperature $T_G$ at which the string network freezes out.
Since $\xi$ depends strongly on temperature near a (second order)
phase transition
--- varying (in mean field theory) like
$\xi(T) \sim \xi_0 [1 - (T/T_c)^2]^{-1/2}$ for $T$ below the
critical temperature $T_c \sim v$ --- it is important to include
this temperature dependence when performing the estimates of
interest. Taking the scalar potential to be $V = \lambda (|\phi|^2
- v^2)^2$, the zero-temperature correlation length, $\xi_0$, is
obtained by equating $\phi$'s derivative energy, $v^2/\xi_0^2$ and
potential energy, $\lambda v^4$, leading to $\xi_0 \sim
\lambda^{-1/2}v^{-1}$. As the temperature falls below $T_c$, $\xi$
shrinks and spatial fluctuations in $\phi$ become increasingly
suppressed. Once $T$ sinks to $T_G$, where $T_c^4 - T_G^4 \sim
\lambda v^4$, then thermal fluctuations are no longer sufficiently
energetic to undo the energy density locked up in vortex domains,
and so according to the Kibble mechanism \cite{kibble}, vortices
freeze in once $T \sim T_G \sim T_c - \lambda T_c$, for which the
correlation length is $\xi_G = \xi(T_G) \sim \lambda^{-1} v^{-1}
\gg \xi_0$.

Once the universe falls below, $T_G \sim T_c \sim v$, we therefore
expect of order $N_{\rm str} \sim (\ell/\xi_G)^2 = 1/(\xi_G H)^2
\sim (\lambda v/H)^2$ such strings in a region whose dimension is
the Hubble scale, $\ell = 1/H$. The tension of each string is of
order $\cT \sim (\lambda v^4) \, \xi_G^2 \sim v^2/\lambda$, and so
the energy carried by such strings in a Hubble volume is
$\rho_{\rm str} \ell^3 \sim N_{\rm str} \cT \ell$, or $\rho_{\rm
str} \sim N_{\rm str} \cT H^2 \sim \lambda v^4$. This is to be
compared with the total energy density at this time, $\rho_{\rm
tot}$, which satisfies $\rho_{\rm tot} \sim T_G^4 \sim T_c^4 \sim
v^4$ and so the energy fraction invested into the cosmic string
network according to this estimate is of order $\rho_{\rm
str}/\rho_{\rm tot} \sim \lambda \ll 1$.

The lifetime of such a string network is similarly set by the rate
with which they chop themselves up into small loops which can then
radiate away their energy. This process is very efficient for
gauge-theory strings, which have roughly 100\% probability of
breaking and reconnecting when two strings cross one another.

What appears to be required, then, is a string network which is
particularly frustrated in comparison with the string networks
which are generated by weakly-coupled gauge models, inasmuch as it
must be more difficult than usual for the strings to chop
themselves up into small loops which can then radiate away their
energy. Interestingly, frustration appears to be a natural feature
of the networks containing D- and F-strings which can arise in
brane-antibrane annihilation. It would be very interesting to
determine whether or not such networks can actually come to
dominate the energy density of the universe for sufficiently long times
after inflation.


\section{A Toy Model}

In this section we seek a simple toy model which has several
separate inflationary epochs. We do so in order to understand the
nature of the fine-tunings which are likely to be required in
order to embed a multi-step inflationary process into a more
sophisticated string-based framework. We first do so for a simple
multiple-scalar model, and then speculate on how such models might
be embedded into the field theories which have been considered
recently within the context of string-motivated inflation.

For these purposes let us consider a simple multiple hybrid
inflationary scenario produced by the following scalar potential
\be
\label{Vpot}
    V=\lambda_{1}^2(\phi^2-\phi_c^2)^2+
    \lambda_{2}^2S^2\phi^2+\frac{m^2}{2}S^2\,
    +g_{1}^2(\chi^2-\chi_{c}^2)^2+g_{2}^2S^2\chi^2 +...\,.
\ee
Such a potential can be motivated in supersymmetric theory, where the
vevs are arising from various D-terms. We may assume more than one
anomalous $U(1)$'s, the anomaly cancellation via the Green Schwarz
mechanism~\cite{GS} determines the Fayet-Illiopolous D-terms.  The
anomaly cancellation can occur at various stages below the string
scale.

With this potential we look only for three bouts of inflation well
separated from epochs of matter domination~\footnote{Previous
considerations of two bouts of inflation were constructed in a
hybrid scenario, see~\cite{Garcia,Anne}, however in these models
second bout of inflation occurs during the second order phase
transition and in some cases with a right choice of parameters the
number of $e$-foldings can be as large as $N_{e}\sim {\cal
O}(10^{6})$~\cite{Anne}.}.


\FIGURE{\epsfig{file=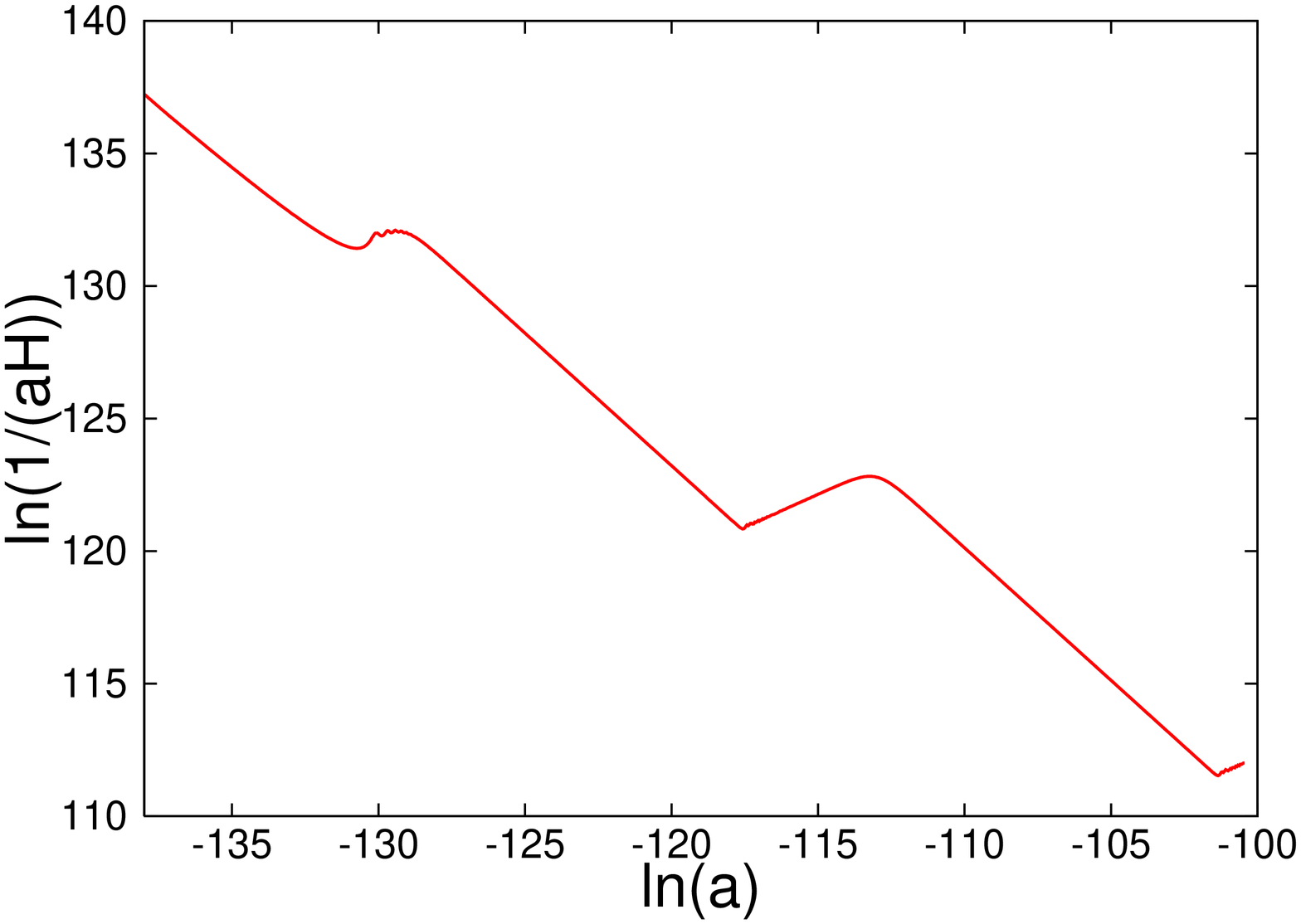,width=8cm,angle=-90}
  \caption{$\ln(a^{-1}H^{-1})$ versus $ln(a)$.}}\label{fig:fig2}

For appropriate choices of the constants the inflaton, $S$, lies along
a flat direction denoted, which slowly rolls down in the potential
obtained when the other fields take approximately constant values,
$\phi\approx 0,\chi\approx 0$, close to their local minimum.  The
dynamics of $S$ field then triggers second order phase transition for
$\phi$ and $\chi$ fields respectively.  Let us assume that the vev of
the inflaton is sufficiently large, $\langle S\rangle \sim M_{\rm p}$,
such that
\begin{equation} \label{phases}
    S> S_{c} = \sqrt{2}
    \left(\frac{\lambda_{1}}{\lambda_{2}}\right)
    \phi_{c}> \sqrt{2}\left(\frac{g_{1}}{g_{2}}
    \right)\chi_{c}\,.
\end{equation}
With this choice, the fields $\phi$ and $\chi$ obtain large masses
from the vev of $S$ and therefore they settle down in their local
minimum given by $\phi=0$ and $\chi=0$. Their dynamics is then
completely decoupled from that of $S$. The potential along the bottom
of the trough parameterized by $S$ is then given approximately by
$V\approx m^2 S^2/2$. For values of $S\geq M_{p}$, the first phase of
inflation is a chaotic type, which comes to an end when $S\sim M_{p}$. The
$S$ field keeps rolling and when $S$ approaches $S_{c}$ the first
phase transition occurs and then subsequently the third.


\FIGURE{\epsfig{file=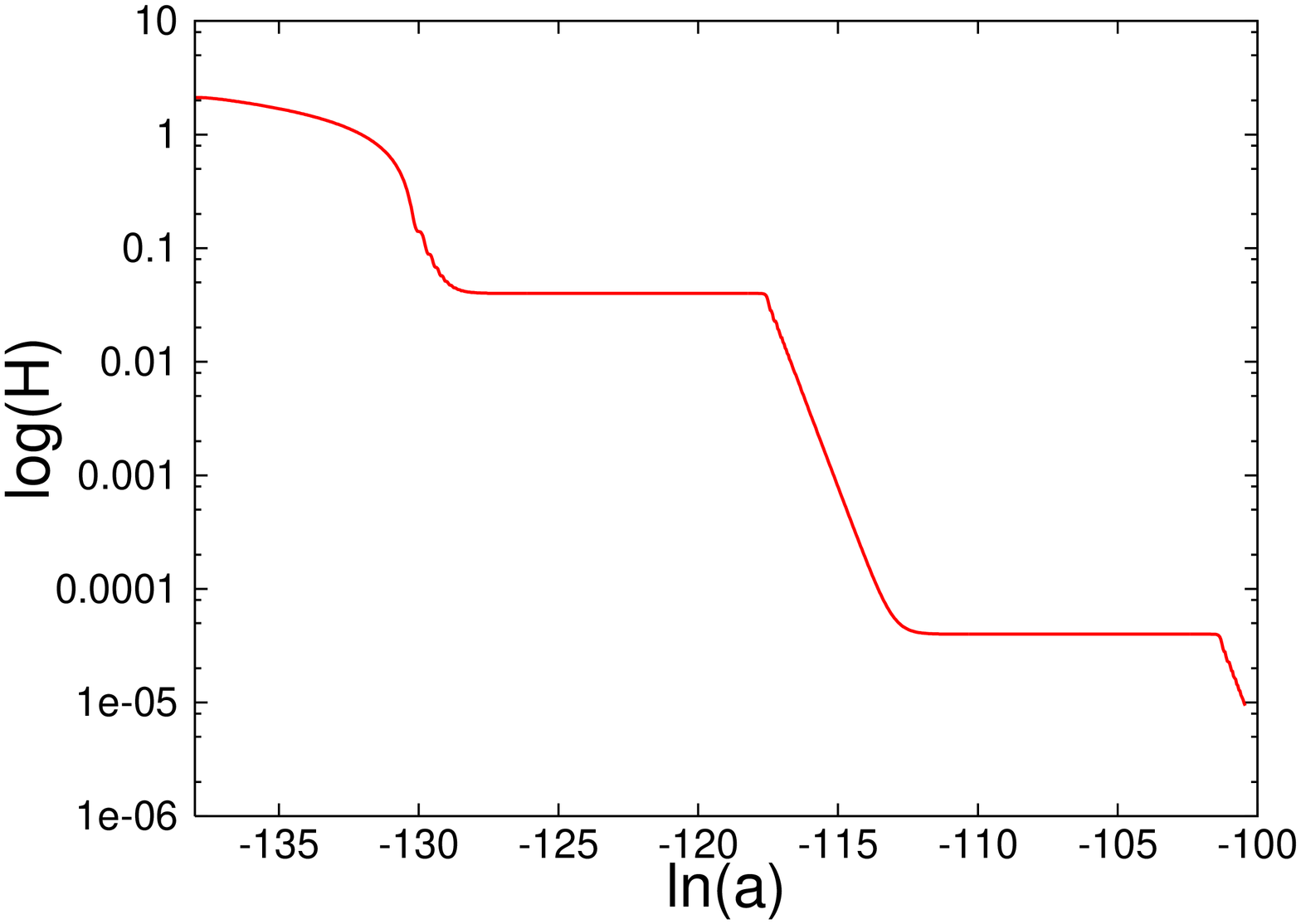,width=8cm,angle=-90}
  \caption{$\ln(H)$ versus $ln(a)$.}}\label{fig:fig3}

After the first phase of inflation the potential is dominated by
the false vacuum of the $\phi$ field,
\begin{equation}
\label{Vinf1}
    V\approx
    \lambda_{1}^2\phi_{c}^4+g_{1}^2\chi_{c}^4+\frac{1}{2}m^2S^2
    \approx \lambda_{1}^2\phi_{c}^4+\frac{1}{2}m^2S^2\,,
\end{equation}
where the latter equality holds when $\lambda_{1}\phi_{c}^2 \gg
g_{1} \chi_{c}^2$.

Our potential is quite rich and depending on the choice of
parameters many interesting possibilities can emerge. For the
purpose of demonstration how the multiple inflation works, we have
studied the evolution of fields in an expanding universe with
these set of parameters and initial conditions; $s_0=3.0 M_{p},\
\phi_0=0,\ \chi_0=0$, $\phi_c=\chi_c=0.2~M_{p}$, $m=1.0~M_{p},\
\lambda_1=1,\ \lambda_2=1,\ g_1=0.001,\ g_2=1$. The vevs and
masses are in Planck units.  The resulting dynamics and expansion
are shown in Figs.(\ref{fig:fig2},\ref{fig:fig3},\ref{fig:fig4}).

For the above choice of parameters the inflaton field slow rolls until
its vev becomes of order the Planck scale. This is the first phase of
inflation. During this period $\ln(H^{-1}/a)$ decreases as shown in
Fig.~(\ref{fig:fig2}). After inflation the $S$ field rolls fast
towards its minimum and on its way down it triggers the dynamics of
$\phi$ and then $\chi$ field. The dynamics of $\chi$ field is
triggered after $\phi$, which can be understood from
eq.~(\ref{phases}), though this is hard to see from the numerical
plots. For a brief period the energy density of the Universe is
dominated by the coherent oscillations of the $S$ field. On average
these oscillations mimic an equation of state that of matter dominated
epoch and the energy density redshifts as $a^{-3}$. During this period
$\ln(H^{-1}/a)$ increases as $(1/3)\ln(t)$, this can be seen in
Fig.~(\ref{fig:fig2}) and in Fig.~(\ref{fig:fig4}). The Hubble
expansion rate drops which is depicted in Fig.~(\ref{fig:fig3}).


\FIGURE{\epsfig{file=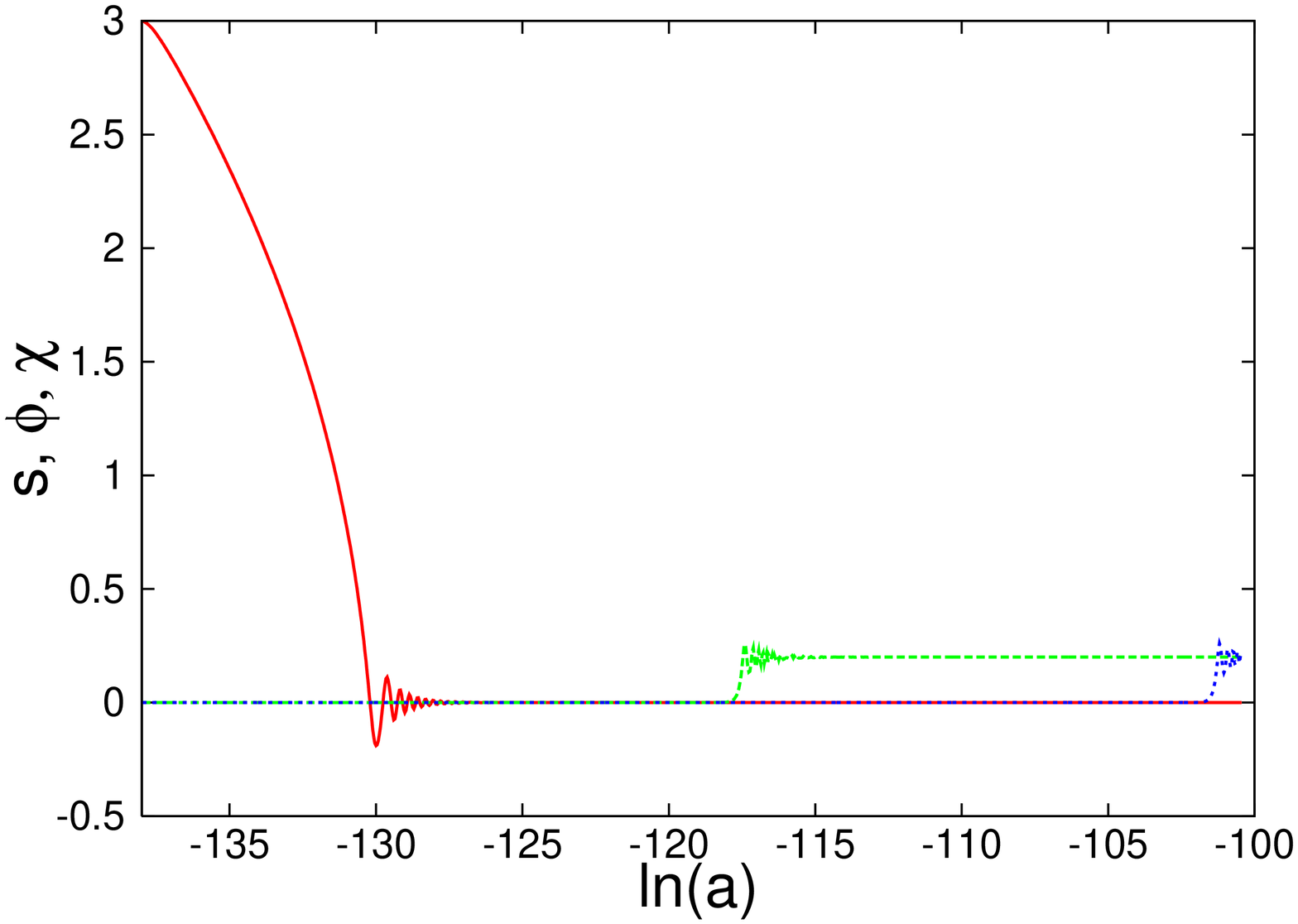,width=8cm,angle=-90}
  \caption{The field
values $s, \phi, \chi$ plotted as function of $ln(a)$.}}\label{fig:fig4}

As the energy density stored in $S$ field redshifts it catches up
with that of the false vacuum of $\phi$ field, and this is the
onset of second bout of inflation.  Note that in between the two
epochs of inflation there is a possibility that some modes which
left the horizon earlier during the first phase can be reprocessed
as they enter the horizon during the second phase of inflation,
see Fig.~(\ref{fig:fig1}). This puts an interesting constraint on
the number of $e$-foldings which we derived in our previous
section. The dynamics of the $\phi$ field is quite interesting,
when the $S$ field started rolling after the end of inflation it
already displace the $\phi$ field from its local maximum due to
presence of the coupling $g_1^2S^2\phi^2$. However $S$ rolls very
slowly towards its global minimum, $\phi_{c}$. In this respect the
second bout of inflation is due to slow rolling of the $\phi$
field, which is similar to the case where the second order phase
transition happens extremely slowly studied earlier in
ref.~\cite{Anne}.

After the second bout of inflation the $\phi$ field oscillates
coherently and its oscillations dominate the energy density of the
Universe before being taken over by the false vacuum energy density of
the $\chi$ field. During the $\phi$ oscillations the Universe enters
briefly into a matter dominated epoch. This is shown in
Fig.~(\ref{fig:fig2}) as a second break in $\ln(H^{-1}/a)$.  The
oscillations of $\phi$ field can be seen in Fig.~(\ref{fig:fig4}).
The dynamical evolution of the Universe after this phase is solely
determined by the potential created by the $\chi$ field. The Universe
enters the third epoch of inflation, which is shown in
Fig.~(\ref{fig:fig2},\ref{fig:fig3}). The comoving Hubble length
decreases with time until $\chi$ begins its oscillations.

For other choice of parameters it is possible to show that the $S$
field keeps rolling while inflation occurs due to the false vacuum
of $\phi$ field and then by $\chi$ field. These two fields
decouple gradually from the dynamics followed by the $S$ field.
The full dynamical analysis of this toy model will be given in a
separate publication. It is a matter of simple extension that our
model can predict several bouts of inflation one after the other
followed by the epochs of matter domination.

It is tempting to try to embed this kind of construction into the
scalar sector of the supergravity theories which have emerged
recently within discussions of string-inspired inflationary
models. This is of particular interest given the potentially
natural occurrence there of string networks at the end of the
first inflationary period.

Perhaps the simplest model of this type to consider is to imagine
having an initial period of brane-antibrane inflation within a
warped throat, perhaps along the lines of explicit models examined
in refs.~\cite{KKetalT,RealInfl}. The success of this inflationary
epoch relies on fairly detailed tuning of the parameters of the
inflaton potential, and this tuning would be less severe if only
the 12 $e$-foldings of inflation found here were required.

Once the initial brane and antibrane have annihilated, their
relative displacement is no longer available to be inflatons
within the low-energy effective theory. A simple way to obtain a
second inflationary phase could then be to use a bulk modulus as
an inflaton, along the lines of the `racetrack' inflationary
mechanism \cite{Racetrack}. This mechanism also requires
fine-tuning of scalar parameters, although we have seen that only
of order 20 $e$-foldings would be required for this second
inflationary phase. It is an open question whether the
fine-tunings required to obtain these two kinds of inflation can
be made compatible with one another.

\section{Conclusions}

In a nutshell, our conclusions are these: A two-step inflationary
process, with an initial inflationary scale of order $V_I^{1/4}
\sim 10^{16}$ GeV and a subsequent scale of order $V_{II}^{1/4}
\sim 10^3$ GeV, can take place without having observable effects
on those fluctuations which are presently observable on
cosmological scales. How much inflation is required during these
two epochs turns out to depend crucially on the equation of state
of the universe during the inter-inflationary phase, as follows.

\begin{itemize}
\item Requiring that the scales which are accessible to
cosmological observations  leave the horizon during the first
inflationary period implies that this period must be at least 7
$e$-foldings long.

\item The requirement that the universe expands enough to dilute
the energy density from $V_I$ to $V_{II}$ ensures that an enormous
amount of expansion occurs during the inter-inflationary phase.
For most non-inflationary equations of state, many of the modes
which leave the horizon by the end of the first inflationary epoch
will re-enter the horizon between the inflationary epochs. For
matter- or radiation-dominated equations of state the requirement
that none of the observable modes get reprocessed in this way
implies that there must be at least about 20 $e$-foldings of
inflation between the exit of the last observable mode and the end
of the first inflationary period.

\item Solving the flatness and horizon problems of the late
universe needs a total amount of inflation in both epochs of at
least about 32 $e$-foldings.  This number increases with the
energy scale, $V_{II}$, of the second inflationary period.

\item In the special case that the equation
of state during the inter-inflationary phase is that of a cosmic
string network, $p = - \rho/3$, the product $aH$ remains constant
and modes neither enter nor leave the horizon during this
phase. In this case modes sufficiently close to the horizon do get
re-processed. In a simple model we find that that no
observable effects for the CMB arise if there are around 5
$e$-foldings of inflation between the exit of the observable modes
and the end of the first inflationary period. In this case only
7+5 = 12 $e$-foldings would be required during the first
inflationary period, with a further 20 being required during the
second inflationary epoch.
\end{itemize}

We have pursued this model as an example of the new and unexpected
options presented to the inflationary model builder by the string
landscape. The present analysis is a first pass at understanding
the phenomenology of this two phase model of inflation, and relies
on a number of simplifying assumptions in order to capture the
basic physics of the model. Interestingly, while the use of two
stages of inflation means that this model is much more complicated
that many other inflationary scenarios, it does not depend on
carefully tuned numerical coefficients. Indeed, the  10  or so
$e$-folds of inflation at high scales is roughly the number one
finds in the {\em absence\/} of tuning for broad classes of
inflationary models. Likewise, the 25 $e$-folds of inflation
needed at the TeV scale for the particular example we consider
here is natural in some contexts, such as locked inflation
\cite{dk,Easther:2004qs}.  Moreover, since the second period of
inflation does not produce the density perturbations we see on the
sky,  we could imagine it being produced by a much broader range
of mechanisms or potentials than we could normally consider.


\section*{Acknowledgments}
We thank Robert Brandenberger and Andrew Liddle for fruitful
discussions. The research of C.B. is supported in part by funds
from NSERC of Canada. R.E. is supported in part by the United
States Department of Energy, grant DE-FG02-92ER-40704. A.M., T.M.
and D.F.M. are partly supported by the Nordic Network Project on
Particle Cosmology. D.F.M acknowledges support from the Research
Council of Norway through grant number 159637/V30.


\section{Appendix}


\FIGURE{\epsfig{file=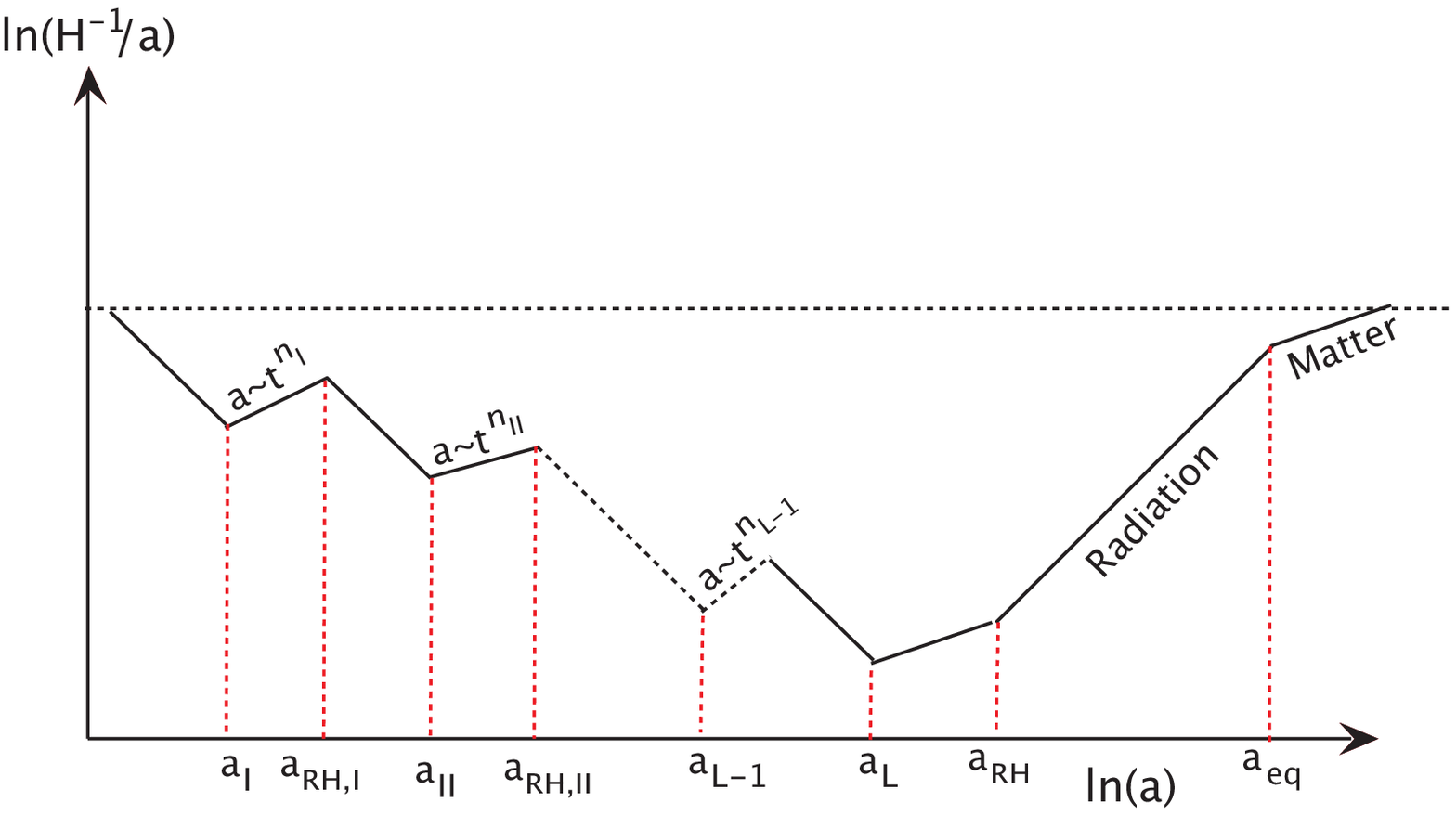,width=12cm}
  \caption{A sketch of the comoving Hubble scale during a multi-step
           inflationary process. We plotted $\ln(H^{-1}/a)$ versus
           $\ln(a)$. We assumed inflationary epochs are followed by
           normal power law expnasions with $n_1,n_2,.., n_L\leq 1$.
           We also assume that after the final bout of inflation
           the Universe is finally settled in a radiation and then in
           a matter dominated epoch. }}\label{fig:apfig1}

We present a scenario where there are mutiple stages of inflation
one after the other separated by the matter epochs, whose
expansions are parameterized by $t^{n_1},~t^{n_2},...,~t^{n_L}$,
with $n_1,~n_2,...,n_L \leq 1$.  We define the $e$-foldings during
the $L$ different inflationary periods as
\be\label{efolddefsap}
  \e^{N_I(k)}\equiv {a_I\over a_k},\ \
  \e^{N_{i}}\equiv {a_{i}\over a_{RH,i-1}},\ i=2,3,...,L\,,
\ee
where $a_{i}$ is the value of the scale factor during the end of the
$i$th phase of inflation and $a_{RH,i}$ during the end of the $i$th
reheating period.

In addition, we define the number of $e$-foldings of expansion
when the critical scale leaves the horizon before the end of first
inflationary period
\be\label{efolddefs2ap}
    \e^{\N}\equiv {a_I\over a_{k_c}}\,.
\ee
It is now straightforward to generalize eq.~(\ref{efolds}):
\bea \label{efoldsap}
     {k\over a_0H_0} & = & {a_kH_k\over a_0H_0}\\
     & = & \Big({a_k\over a_I}\Big)\Big({a_I\over a_{RH,I}}\Big)
           \Big({a_{RH,I}\over a_{II}}\Big)
           \Big({a_{II}\over a_{RH,II}}\Big)...
           \Big({a_{RH,L-1}\over a_{L}}\Big)
           \Big({a_{L}\over a_{RH}}\Big)
           \Big({a_{RH}\over a_{eq}}\Big)
           \Big({H_k\over H_{eq}}\Big)
           \Big({a_{eq}H_{eq}\over a_0H_0}\Big)\nonumber\\
    & = &  \e^{-N_I(k)}\Big({a_I\over a_{RH,I}}\Big)\e^{-N_{II}}
           \Big({a_{II}\over a_{RH,II}}\Big)...
           e^{-N_L}\Big({\rho_{L}\over \rho_{RH}}\Big)^{-1/3}
           \Big({\rho_{RH,II}\over \rho_{eq}}\Big)^{-1/4}
           \Big({H_k\over H_{eq}}\Big)
           \Big({a_{eq}H_{eq}\over a_0H_0}\Big)\,.\nonumber
\eea
Since during the period between $i$ and $i+1$ phases of inflation, the
Universe expands as $a\sim t^n_i$.  Hence during these periods,
$\ln(1/(aH))\sim (1-n_i)/n_i \ln(a)$, and therefore
\be\label{infcalcapp}
        \ln\Big({a_i\over a_{RH,i}}\Big)={n_i\over 2}
                  \ln\Big({V_{i+1}\over V_i}\Big)\,.
\ee
Substituting eq.~(\ref{infcalcapp}) into eq.~(\ref{efoldsap}),
and a careful rearrangement gives
\bea\label{lnefoldsapp}
        N_I(k)+\sum_{i=2}^L N_i & = & -\ln\Big({k\over a_0H_0}\Big)+
        \frac 13 \ln\Big({\rho_{RH,L}\over \rho_{L}}\Big)+
        \frac 14 \ln\Big({\rho_{eq}\over \rho_{RH,L}}\Big)\nonumber\\
        & & +\ln\Big({\sqrt{8\pi V_k\over 3 m_{Pl}^2}{1\over H_{eq}}}\Big)+
        \ln(219\, \Omega_0h)+\sum_{i=1}^{L-1}{n_i\over 2}
        \ln\Big({V_{i+1}\over V_i}\Big)\,.
\eea
We can assume instant reheating after the last phase of inflation,
which yields, $\rho_{L,end}=\rho_{RH,L}$.  Concentrating on the
present horizon scale (we define $N_I\equiv N_I(a_0H_0)$), we find
\be\label{lnefolds2app}
        \sum_{i=1}^L N_i=68.5+\frac
        14\ln\Big({V^2_{I}\over V_{L}m_{Pl}^4}\Big)+
        \sum_{i=1}^{L-1}{n_i\over 2}\ln\Big({V_{i+1}\over V_i}\Big)\,.
\ee
In comparison to the two-stage inflationary epochs, there is now a
set of constraints arising from the fact that there should be no
reprocessing of modes on the scales probed by CMB experiments.
This requirement constraints
\be\label{noreprosapp}
         \ln\Big({1\over a_{k_c}H_I}\Big)>
         \ln\Big({1\over a_{RH,i}H_{i+1}}\Big)\,,
         \ \ i=2,...,L-1\,
\ee
or,
\be\label{norepros2app}
          \ln\Big({a_{RH,i}\over a_{k_c}}\Big)>\frac 12
          \ln\Big({V_I\over V_{i+1}}\Big)\,,\ \ i=2,...,L-1\,.
\ee
The left hand side can be expanded:
\bea\label{norepros3app}
          \ln\Big({a_{RH,i}\over a_{k_c}}\Big) & = &
          \ln\Big({a_{I}\over a_{k_c}}{a_{RH,I}\over a_{I}}{a_{II}
          \over a_{RH,I}}...{a_{RH,i}\over a_{i}}\Big)\nonumber\\
          & = & \N-{n_1\over 2}\ln\Big({V_{II}\over V_I}\Big)+N_{II}
          -{n_2\over 2}\ln\Big({V_{III}\over V_{II}}\Big)+...+N_i-
          {n_i\over 2}\ln\Big({V_{i+1}\over V_i}\Big)\nonumber\\
          & = & \N+\sum_{j=2}^{i}N_j-\sum_{j=1}^{i}{n_j\over 2}
          \ln\Big({V_{i+1}\over V_i}\Big)\nonumber\\
          & = & \frac 12 \ln\Big({V_{I}\over V_{II}}\Big)+
          \sum_{j=2}^{i}N_j+\sum_{j=2}^{i}{n_j\over 2}
          \ln\Big({V_{i}\over V_{i+1}}\Big)\,,
\eea
where in the last step we have used $\N=(1-n_1)/2\ln(V_I/V_{II})$.

Hence, in general we obtain the following set of constraints:
\bea\label{genconstsapp}
         & & N_I>6.9+{1-n_1\over 2}\ln\Big({V_I\over V_{II}}\Big)
         \nonumber\\ & & \sum_{j=2}^{i}N_j  >
         \frac 12 \ln\Big({V_{II}\over V_{i+1}}\Big)
         -\sum_{j=2}^{i}{n_j\over 2}\ln\Big({V_{i}\over V_{i+1}}\Big)\,,
         \ \ i=2,...,L-1\,.
\eea
In the special case $n_i=1$, it is easy to see that these constraints are
trivially satisfied ($\sum_jN_j>0$).



\end{document}